\documentclass{optica-article-pr2col}

\journal{prj}

\articletype{Research Article}

\usepackage{epstopdf}
\newcommand{\prfrontmatter}{%
\title{MCP-based EUV ptychography with physics-simulation-trained residual learning}
\author{Qijun You,\authormark{1} Nuoxuan Yang,\authormark{1} Fangrui Quan,\authormark{1} Xu Han,\authormark{2} Jie Li,\authormark{3} Seunghwoi Han,\authormark{4} and Wei Cao\authormark{1,*}}
\address{\authormark{1}School of Physics and Wuhan National Laboratory for Optoelectronics, Huazhong University of Science and Technology, Wuhan 430074, China\\
\authormark{2}North Night Vision Technology Co., Ltd., No. 5, Hongwai Road, Kunming 650106, Yunnan, China\\
\authormark{3}Aerospace Information Research Institute, Chinese Academy of Sciences, Beijing, 100094, China\\
\authormark{4}Ultra Precision Photonics Laboratory, Department of Mechanical Engineering, Chonnam National University, Gwangju, 61186, South Korea}
\email{\authormark{*}weicao@hust.edu.cn}
\begin{abstract*}
Microchannel-plate (MCP) detectors offer high gain for extreme-ultraviolet (EUV) experiments, but spatial blur and nonlinear response hinder quantitative diffraction imaging. Here, we present an MCP-based EUV ptychography system that combines detector hardware with physics-simulation-trained computational restoration. A residual-learning neural network trained on simulated detector degradation restores diffraction patterns without experimentally acquired training pairs. This preprocessing substantially improves image quality compared with reconstructions from uncorrected data under identical reconstruction settings. Single-harmonic imaging at $57$~nm yields an effective edge-response width of $1.08~\mu$m, approaching the nominal diffraction-limited scale of the imaging geometry. The approach further enables wavelength-multiplexed imaging at $44.4$ and $57$~nm. By coupling high-gain detection with computational compensation, this system enables high-resolution, multi-harmonic EUV ptychography and offers a cost-effective alternative to scientific-grade EUV CCD detectors.
\end{abstract*}
\vspace{1.5\baselineskip}
}

\begin{document}

\twocolumn[\prfrontmatter]
\sloppy

\section{Introduction}

Coherent diffractive imaging (CDI) retrieves a sample's complex transmission from measured diffraction intensities without an image-forming lens \cite{ph2}. In ptychography, diffraction patterns are recorded from overlapping illuminated regions at different sample positions. This redundancy constrains the joint recovery of the object and illumination probe and improves reconstruction robustness \cite{ptychography1,epie}. High-order harmonic generation (HHG) brings coherent extreme-ultraviolet (EUV) illumination to laboratory-scale systems, enabling short-wavelength imaging, wavefront characterization, and spectroscopy \cite{broadband5,bio1,Du23}. These applications require sensitive detectors that capture weak diffraction signals while preserving the spatial information needed for phase retrieval.

\begin{figure*}
	\includegraphics[width=\linewidth]{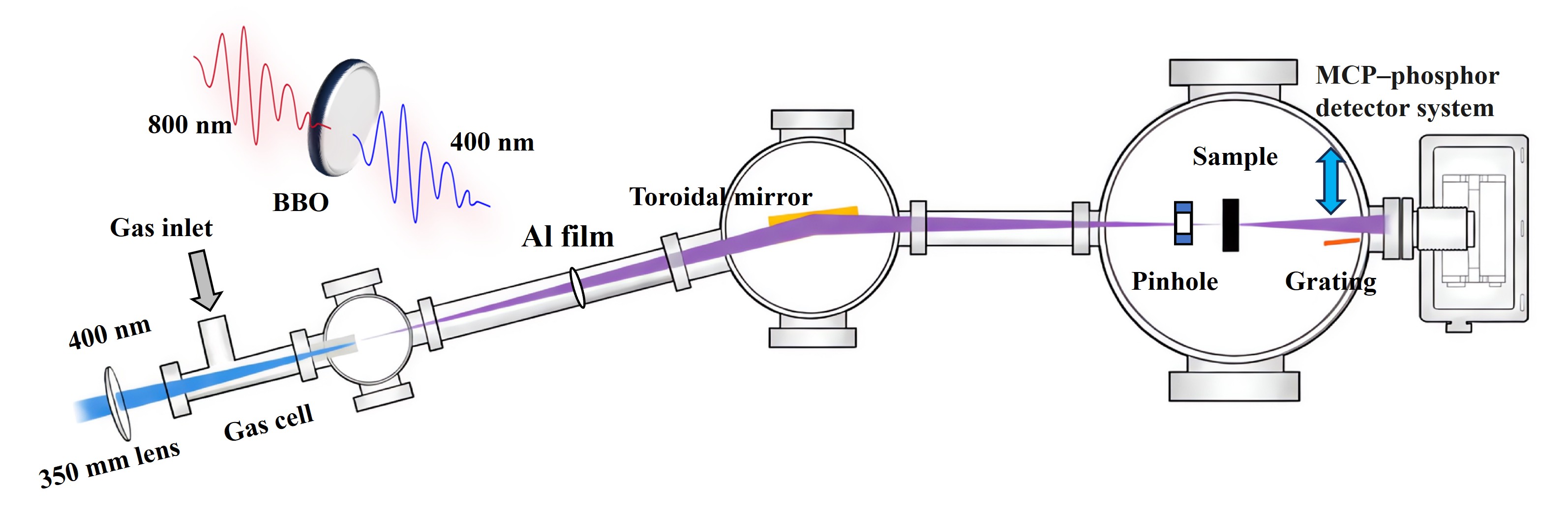}%
	\caption{\label{fig:apparatus} Schematic of the extreme-ultraviolet ptychographic imaging apparatus. The fundamental $800$~nm pulses from a Ti:sapphire laser are frequency-doubled in a $\beta$-barium borate (BBO) crystal, focused into an argon gas cell to generate high-order harmonics, filtered by an aluminum film to remove photons below $15$~eV, and refocused by a toroidal mirror ($2f=500$~mm) into the imaging-and-spectrometer chamber. The chamber supports both coherent diffraction imaging with an MCP--phosphor detector and grating-based harmonic spectroscopy. The differential pumping scheme keeps the imaging chamber below $4\times10^{-5}$~Pa.}
\end{figure*}

Scientific-grade direct-detection EUV cameras are widely used in HHG-based imaging \cite{Brooks2022}, but specialized sensor processing and vacuum-compatible packaging contribute to their high cost. Microchannel-plate (MCP) detectors offer an alternative by converting incident radiation into amplified electron signals \cite{Wiza1979}. Coupled to a phosphor screen and a commercial visible camera, they combine high gain, a large active area, and flexible optical readout. MCP-based detection has enabled single-shot HHG diffraction measurements of isolated helium nanodroplets, with particle properties extracted using multicolor scattering models \cite{Rupp2017}. Extending MCP detection to iterative ptychographic imaging would broaden access to coherent EUV measurements, provided that detector-induced distortions can be adequately compensated.

\begin{figure}
	\includegraphics[width=\linewidth]{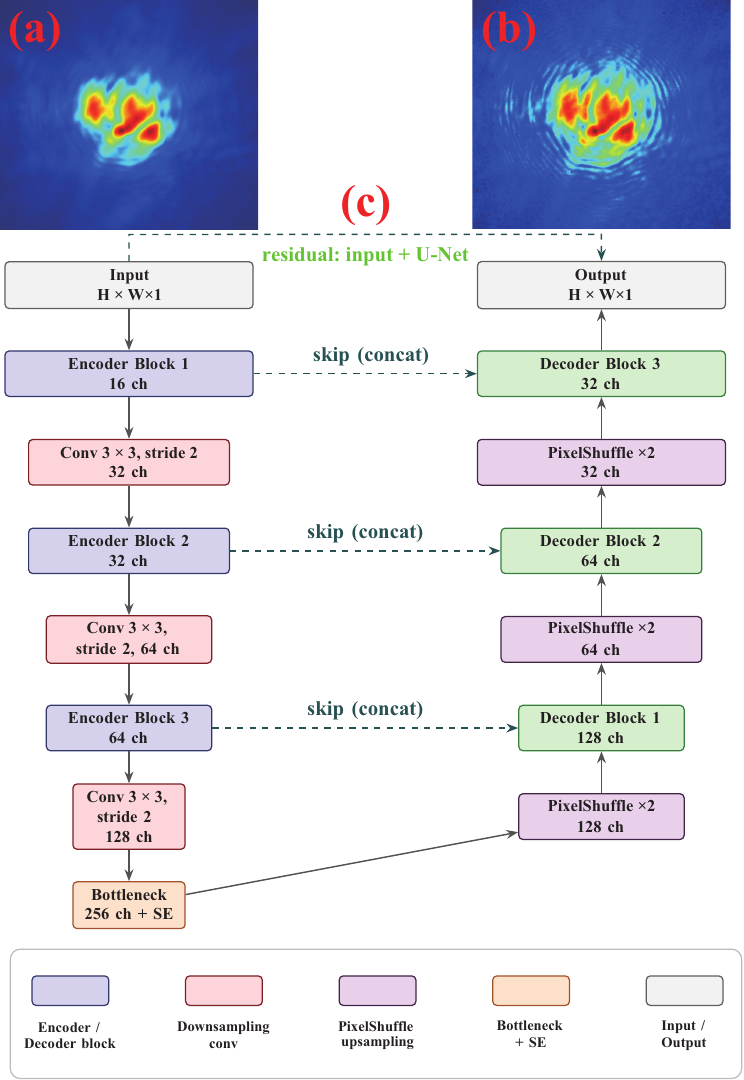}%
	\caption{\label{fig:deblur} CNN deblurring of MCP-degraded diffraction patterns. (a) Raw (degraded) diffraction pattern. (b) Corrected (deblurred) high-quality diffraction pattern. (c) Architecture of the residual-learning deblurring network.}
\end{figure}

The principal challenge is that high MCP gain is accompanied by spatial spreading and intensity-dependent degradation. Electron-cloud expansion between successive plates and across the MCP-to-phosphor gap broadens the detector response beyond the individual channel diameter \cite{Wiza1979,HamamatsuMCP}. Shortening the proximity gap can reduce the latter contribution but does not eliminate spreading within the cascade. Gain fluctuations, nonlinear response, and noise further distort the recorded intensities. Together, these effects reduce diffraction-fringe visibility and introduce a mismatch between measured data and the ideal-detector model used in phase retrieval. Hardware optimization alone therefore cannot fully resolve the trade-off between high gain and fine spatial response. Computational compensation offers a complementary route to improving imaging performance while retaining cost-effective detection hardware.

Recent advances in deep learning have enabled data-driven signal restoration in electron diffractive imaging and X-ray ptychography \cite{Chang2023,Li2026}. For MCP-based EUV experiments, however, acquiring paired degraded and clean experimental diffraction patterns for supervised training is challenging. A physics-based simulation of the detector chain can instead generate paired training data by applying spatial blur, stochastic gain and noise, and nonlinear response to simulated diffraction patterns. Residual learning \cite{He2016} with a U-Net architecture \cite{Ronneberger2015} can then estimate corrections to the degraded intensities without experimentally acquired training pairs. Unlike enhancement of the final reconstructed image, this strategy addresses detector-induced distortions before phase retrieval, aiming to improve the consistency of the diffraction data with the reconstruction model.

Here, we present an MCP-based EUV ptychography system that integrates high-gain detection, physics-simulation-trained restoration, and iterative reconstruction. Comparisons using the same single-harmonic dataset and identical reconstruction settings demonstrate substantially improved imaging quality after restoration. Imaging at $57$~nm yields an effective edge-response width of $1.08~\mu$m, approaching the nominal diffraction-limited scale of the imaging geometry. Combined with ptychographic information multiplexing (PIM) \cite{broadband2,broadband3}, the approach further enables simultaneous wavelength-resolved sample imaging and probe-wavefront retrieval at $44.4$ and $57$~nm. By addressing detector degradation computationally rather than relying solely on hardware refinement, this work enables high-resolution, multi-harmonic EUV ptychography and offers a cost-effective alternative to scientific-grade EUV CCD detectors.

\section{Apparatus}

A schematic of the compact EUV ptychographic imaging apparatus is shown in Fig.~\ref{fig:apparatus}. The system consists of a laser and frequency-doubling module, a high-order harmonic generation module, a filtering and refocusing module, and an imaging-and-spectrometer chamber.

A Ti:sapphire laser delivers $35$~fs pulses centered at $800$~nm at a repetition rate of $1$~kHz. Frequency doubling in a BBO crystal produces $400$~nm pulses, which are focused into an argon gas cell to generate high-order harmonics \cite{Corkum1993}. The seventh and ninth orders, at approximately $57$~nm ($21.7$~eV) and $44.4$~nm ($27.9$~eV), respectively, are used for imaging.

A $500$-nm-thick aluminum filter suppresses radiation below approximately $15$~eV, and a gold-coated toroidal mirror ($2f=500$~mm, $7^\circ$ grazing incidence) refocuses the transmitted beam. A circular pinhole placed approximately $2$~cm upstream of the sample defines the illumination spot, while the sample is mounted on a two-dimensional piezoelectric nanopositioning stage. Far-field diffraction patterns are recorded by a chevron-stacked MCP–phosphor detector located approximately $36$~cm downstream of the sample. By translating the sample out of the beam and inserting a flat-field grating mounted on a single-axis stage, the same MCP assembly records the dispersed harmonic spectrum, eliminating the need for a dedicated EUV spectrometer detector.

Because high-voltage MCP operation requires a background pressure below $4\times10^{-5}$~Pa, the argon gas cell is sealed except for a single laser-drilled micro-aperture that acts as both the optical entrance and the sole gas conductance path. This differential-pumping geometry maintains the imaging chamber below $4\times10^{-5}$~Pa under a $3000$~Pa gas-cell load, suppressing residual-gas ion feedback and background glow on the MCP. For driving-pulse energies of $0.2$--$0.4$~mJ, the single-pulse energy delivered to the sample is estimated to be on the order of $20$~nJ.

The detector assembly consists of a chevron stack of two MCPs with $8$~$\mu$m pore diameter (North Night Vision Technology Co., Ltd., Kunming, China), proximity-focused onto a P43 ($\mathrm{Gd_2O_2S{:}Tb}$) phosphor screen. The phosphor image is relayed by a $0.59\times$ demagnifying lens onto a $2048\times2048$ visible CMOS camera with $6.5$~$\mu$m pixels, corresponding to an effective sampling pitch of $11.0$~$\mu$m per pixel at the phosphor plane and a $22.6\times22.6$~mm$^2$ active field of view. According to manufacturer specifications, the spatial point-spread function (PSF) scales approximately with the channel diameter: a single $8$~$\mu$m-pore plate exhibits a full width at half maximum (FWHM) of $\sim30$~$\mu$m, while a chevron stack broadens the response to roughly $55$--$65$~$\mu$m. Although we shortened the MCP-to-phosphor proximity gap to reduce post-plate electron divergence, inter-plate charge-cloud spreading remains a hardware limit. Smaller pores ($\le6$~$\mu$m) would narrow the PSF proportionally but reduce the open-area ratio and increase fabrication cost; the $8$~$\mu$m pore geometry was therefore selected as a practical balance among spatial resolution, sensitivity, and subsystem cost (typically an order of magnitude lower than a comparable large-area back-illuminated EUV CCD).

\section{Physics-Simulation-Trained Residual Learning}

Even with the hardware optimization described above, the residual blur still prevents the stitched multi-exposure diffraction data from being used directly for ptychographic reconstruction. Prior HHG-based CDI experiments have corrected only the saturation nonlinearity of the detector (for instance, through an exponential response relating the detected and true intensities \cite{Rupp2017}), leaving the spatial blur caused by electron-cloud spreading uncorrected. To restore the spatial-frequency fidelity required for ptychography, we construct a physics-based forward model of the MCP--phosphor--camera degradation chain to synthesize paired training data, and train a residual U-Net to invert these degradations prior to phase retrieval (Fig.~\ref{fig:deblur}).
\subsection{Physics-based degradation simulation}

The complete degradation chain of the MCP--phosphor detector is modeled as follows. The incident signal is first attenuated by the open-area ratio of the MCP and converted into an expected number of photoelectrons, whose Poisson sampling gives the shot noise. Each electron is then multiplied by the chevron gain. A single MCP has an approximately exponential pulse-height distribution, whereas a chevron stack operated in saturation produces a narrower, quasi-Gaussian distribution. We approximate the pulse-height gain for \(n\) electrons by \(\mathcal{G}(n)=G\,\mathrm{Gam}(n,1)\), where \(G\) is the mean gain per electron and \(\mathrm{Gam}(n,1)\) is a gamma variable with shape \(n\) and unit scale, whose mean and variance are both \(n\). Equivalently, \(n\) independent exponential pulses of mean \(G\) sum to this distribution, so the total charge has mean \(nG\) and relative fluctuation \(1/\sqrt{n}\). For \(n=1\) this reduces to the exponential single-electron pulse-height distribution, and for large \(n\) the aggregate approaches a Gaussian. Because a saturated chevron has a narrower single-electron pulse-height distribution, this choice overestimates the gain fluctuation and is therefore conservative. The relative fluctuation is of order unity for the few-electron events that dominate the weak outer fringes. Poisson dark counts from thermal emission and ion feedback are added before the gain, so that they are amplified in the same way. The charge is then multiplied by a fixed-pattern gain field, softly saturated at the double-MCP output, and passed through the P43 phosphor response. It is finally convolved with the PSF that describes electron-cloud spreading and phosphor scattering. The whole chain is expressed as
\begin{equation}
y(\mathbf{r}) = \mathrm{PSF}(\mathbf{r}) \ast \Phi\!\Bigl\{ \mathcal{G}\!\Bigl[ \mathcal{P}\bigl(\eta\, s(\mathbf{r})\bigr) + \mathcal{P}\bigl(n_{\mathrm{dark}}\bigr) \Bigr] \Bigr\},
\label{eq:degradation}
\end{equation}
where $s(\mathbf{r})$ is the ideal diffraction intensity distribution, $\eta$ is the open-area ratio (transmittance) of the MCP, $\mathcal{P}(\cdot)$ denotes Poisson sampling (shot noise), $\mathcal{G}(\cdot)$ denotes the pulse-height gain process of the chevron stack (mean gain $G$), $n_{\mathrm{dark}}$ is the mean dark-count level arising from thermal emission and ion feedback, $\Phi(\cdot)$ denotes the P43 phosphor response, $\mathrm{PSF}(\mathbf{r})$ is the point-spread function corresponding to electron-cloud spreading and phosphor scattering, $\ast$ denotes the convolution operation, and $y(\mathbf{r})$ is the degraded diffraction pattern. The fixed-pattern gain field and the soft saturation act as multiplicative and compressive stages, respectively, and are omitted from Eq.~(\ref{eq:degradation}) for compactness.

The parameters of Eq.~(\ref{eq:degradation}) follow the detector hardware. The open-area ratio is $0.60$, corresponding to $8$~$\mu$m channels on a $10$~$\mu$m pitch. The mean chevron gain is $1\times10^{5}$, consistent with a $1200$~V stack bias. Dark counts are drawn log-uniformly over one decade. The fixed-pattern gain is the product of a channel-scale term and a large-scale term, representing channel-to-channel gain differences and large-scale phosphor-screen and relay-lens nonuniformity, respectively. The phosphor exponent was set to unity, i.e., a linear P43 response. Saturation is applied well above the level reached by the scattered signal, so that it truncates only the direct beam. After convolution the image is divided by the expected gain factor. The mean of the image then equals the blurred ideal intensity, so that the degraded and clean patterns remain in the same units.
At the phosphor plane the sampling is $11.0$~$\mu$m per pixel, obtained by dividing the $6.5$~$\mu$m camera pixel by the $0.59$ relay magnification. The composite PSF combines five contributions. The $8$~$\mu$m channel aperture and the inter-plate charge-cloud spread of the chevron stack, which spans roughly three channel diameters, determine the broadening at the plates. The MCP--phosphor gap spread contributes comparably, so both mechanisms are major sources of blur. The gap contribution is evaluated from the near-focus relation $r=2d\sqrt{E_{\perp}/eV_{\mathrm{acc}}}$ with the $0.5$~mm gap, the $4.6$~kV acceleration voltage, and the $1$~eV transverse emission energy, and is augmented by a Lorentzian tail for elastically scattered electrons. The $4$~$\mu$m P43 grain scattering and the modulation transfer function of the relay lens add the remaining broadening. The five kernels are multiplied in the Fourier domain, and the composite is evaluated on a grid several times wider than the PSF so that the Lorentzian tail is not truncated.

\subsection{Residual-learning deblurring network}
\label{sec:deblur-net}

As illustrated in Fig.~\ref{fig:deblur}, the deblurring network adopts a residual-learning structure \cite{He2016} in which the output is the sum of the input and the residual predicted by a U-Net \cite{Ronneberger2015}, i.e.,
\begin{equation}
\hat{s}(\mathbf{r}) = y(\mathbf{r}) + \mathcal{U}\!\bigl[y(\mathbf{r})\bigr],
\label{eq:unet}
\end{equation}
where $y(\mathbf{r})$ is the input degraded diffraction pattern, $\mathcal{U}(\cdot)$ is the residual learned by the U-Net, and $\hat{s}(\mathbf{r})$ is the corrected high-quality diffraction pattern. The global skip connection forces the network to learn the degradation residual rather than the direct mapping, which makes the learning problem substantially easier. As shown in Fig.~\ref{fig:deblur}(c), the U-Net consists of an encoding path of 16-, 32-, and 64-channel convolutional residual blocks, a 256-channel bottleneck equipped with a channel-attention module, and a decoding path of 128-, 64-, and 32-channel blocks. Adjacent decoder stages are connected by pixel-shuffle upsampling, which avoids the checkerboard artifacts associated with transposed convolution. Skip connections fuse the encoder features at the same scale into the decoder, compensating for the high-frequency detail lost during downsampling.

Training data were generated entirely by simulation, so no experimentally acquired input--output pairs are required. Clean diffraction amplitudes were synthesized from two object types in equal proportion. Binary lithographic masks contain lines, rectangles, and circular holes with randomized feature sizes, absorber transmission, and absorber phase. Histology images provide irregular edges and texture. Mask edges are rendered with error-function transitions, because a hard raster edge would produce comb-like artifacts in the far field. For the histology images, the amplitude and the phase are taken from two different images. The object was illuminated by a randomly sized Gaussian spot, restricted by a circular aperture in most cases. The field transmitted by the object was then propagated from the sample plane to the detector plane by angular-spectrum propagation.

Each clean pattern was degraded at the native detector grid according to Eq.~(\ref{eq:degradation}) and used directly as the network input, with the corresponding clean pattern as the target. The dose (a scalar scaling of $s$) and the dark-count level were drawn log-uniformly over wide ranges. Because the direct-beam spike is two to three orders of magnitude brighter than the scattered speckles, the dose reference was a high percentile of the pattern rather than its maximum. The direct beam was allowed to saturate softly, as in the experiment. Fixed-pattern gain nonuniformity was represented by a small set of fixed fields applied cyclically, so that the network learns to remove an arbitrary fixed pattern rather than one particular pattern.

Degraded and clean images were normalized with a percentile-based scale taken from the degraded image and shared with the clean target. They were then compressed by an inverse hyperbolic sine (asinh) transform. The asinh transform is logarithmic at large values and linear near zero. A diffraction pattern spans several orders of magnitude between the direct beam and the outer speckles. The transform retains the weak high-frequency fringes that a linear scale would lose, while the bright core remains bounded. Taking the scale from a high percentile rather than from the maximum makes the normalization insensitive to the exposure and to isolated hot pixels. Sharing the scale between input and target keeps the two images in the same units, so that the network learns only the degradation residual. Figures~\ref{fig:deblur}(a) and \ref{fig:deblur}(b) show a representative raw diffraction pattern and the corresponding deblurred pattern, respectively.

\begin{figure*}[t]
	\centering
	\includegraphics[width=\linewidth]{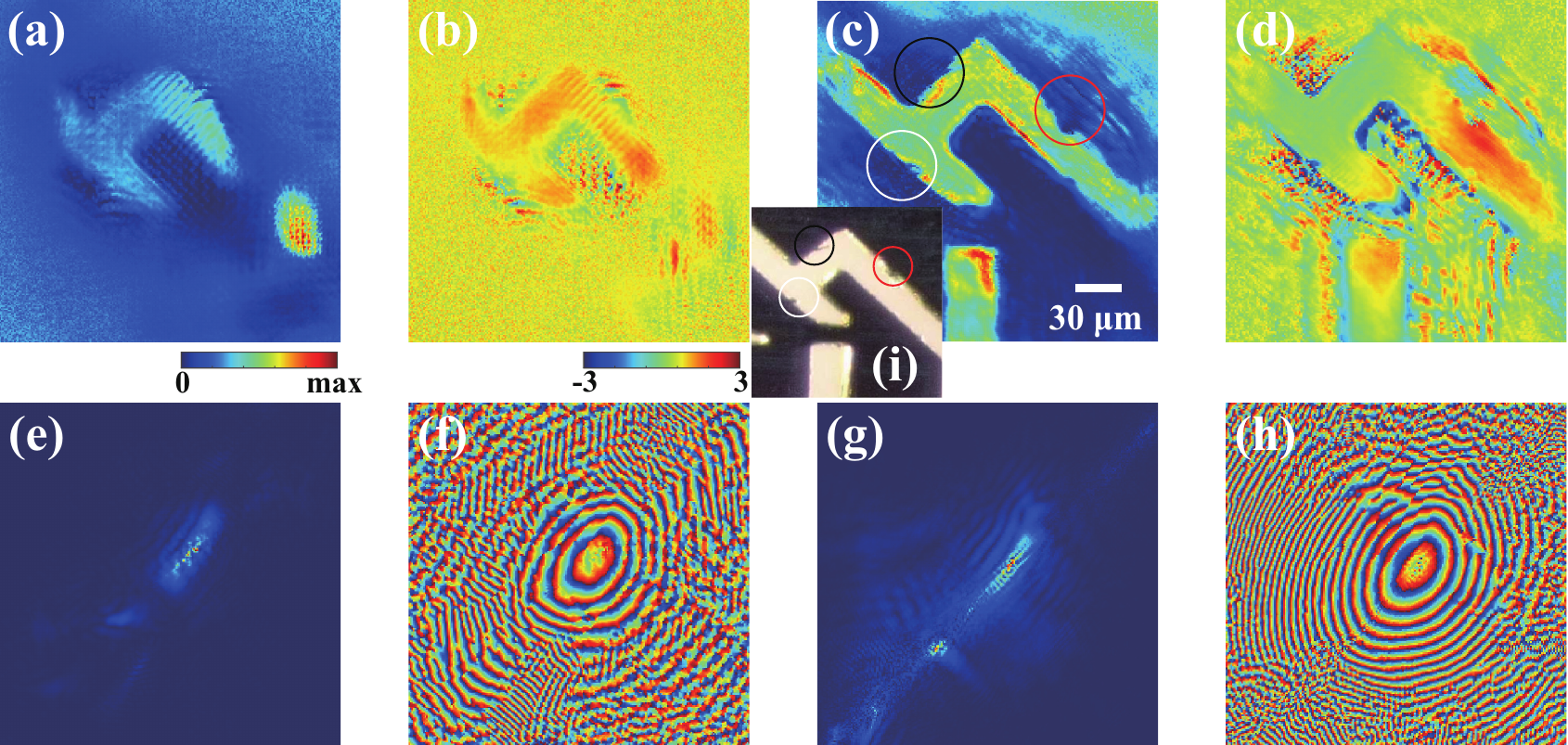}%
	\caption{\label{fig:reconstruction} Single-harmonic ($q=7$, $57$~nm) reconstruction with and without deblurring. (a),(b) Object amplitude and phase reconstructed from the uncorrected diffraction patterns. (c),(d) Object amplitude and phase after deblurring. (e),(f) Probe amplitude and phase reconstructed from the uncorrected diffraction patterns. (g),(h) Probe amplitude and phase after deblurring. (i) Metallographic-microscope image of the same sample; several damaged small features are circled in (c) and (i) to indicate their correspondence.}
\end{figure*}

\section{Ptychographic Reconstruction}

After correction by the trained deblurring network, the high-quality diffraction data are processed by the ptychographic information multiplexing algorithm \cite{broadband2,broadband3}. The data are acquired by scanning the sample along a zigzag path over a nominal $10\times10$ grid ($100$ positions). The stage moves by a nominal step at each point, but the actual step varies within a small interval around this value (about $\pm1$~$\mu$m in the present experiments). The scan points therefore do not form a strictly periodic grid, which suppresses the grid-like artifacts caused by regular sampling. The deviation of every move is recorded, and the scan positions $\mathbf{r}_j$ are determined from the nominal step plus the recorded deviation rather than from an ideal uniform grid. This prevents position mismatch from introducing additional artifacts into the reconstruction.

In PIM the illumination is represented by mutually incoherent channels, and the intensity recorded at each scan position is the incoherent sum of their contributions. The channels are indexed along two independent dimensions. The wavelength mode $w$ labels the harmonic components, which are generated as separate spectral components and are therefore mutually incoherent. The spatial mode $m$ describes the partial coherence of the illumination at a single wavelength, which originates from pointing and energy fluctuations of the source. A probe $P_{m,w}(\mathbf{r})$ and an object $O_{m,w}(\mathbf{r})$ are retrieved for every channel, and the measured intensity is distributed among the channels in proportion to their current contributions. In the single-harmonic experiment the decomposition is applied along the spatial dimension, with one wavelength mode and six spatial modes. In the two-harmonic experiment the seventh and ninth harmonics form the wavelength modes. The complete formulation of the multiplexed reconstruction, including the exit-wave model, the propagation to the detector plane, the modulus constraint, and the update rules, is given in \ref{app:pty}; these steps follow previous work.

Separating closely spaced harmonics from a single diffraction pattern is inherently ill-posed. The contributions of adjacent harmonics could not be unambiguously distinguished even in forward-model fits of single-shot diffraction data \cite{Rupp2017}. The overlapping-scan redundancy of ptychography supplies the additional constraints required for this separation. Multi-wavelength ptychography with a tabletop high-order-harmonic source has demonstrated this separation without monochromatization \cite{Li2024}.

Three additional procedures are applied after each iteration. The probe wavefront can be measured directly with the sample removed, and the summed far-field intensity of all channels is constrained to this measured amplitude. The spatial modes belonging to the same wavelength drift out of orthogonality during the iteration and are re-orthogonalized through the eigendecomposition of their Gram matrix. The orthogonalization is carried out separately for each wavelength group, because fields of different wavelengths are mutually incoherent and their mutual overlap has no physical meaning. Isolated spikes in the probe modes, produced where the update divides by a nearly vanishing illumination, are suppressed by replacing values that exceed twice the local median with that median.

Multiplexing has also been implemented using axial redundancy instead of transverse scanning. Multi-distance lensless imaging separates the spectral responses of a sample from axial diffraction data \cite{You2024}, and single-axis scanning CDI translates the sample along the optical axis for wavelength-resolved reconstruction and wavefront retrieval \cite{You2026}. Spectral modulation has been used to reduce crosstalk between wavelength channels in multiplexed ptychography of broadband pulses \cite{Wang2025}. Reflection ptychography has combined interface and spectral multiplexing to retrieve wavelength-dependent surface structure and topography \cite{Gao2025}.

\section{Experimental Results}

\begin{figure}
	\centering
	\includegraphics[width=\linewidth]{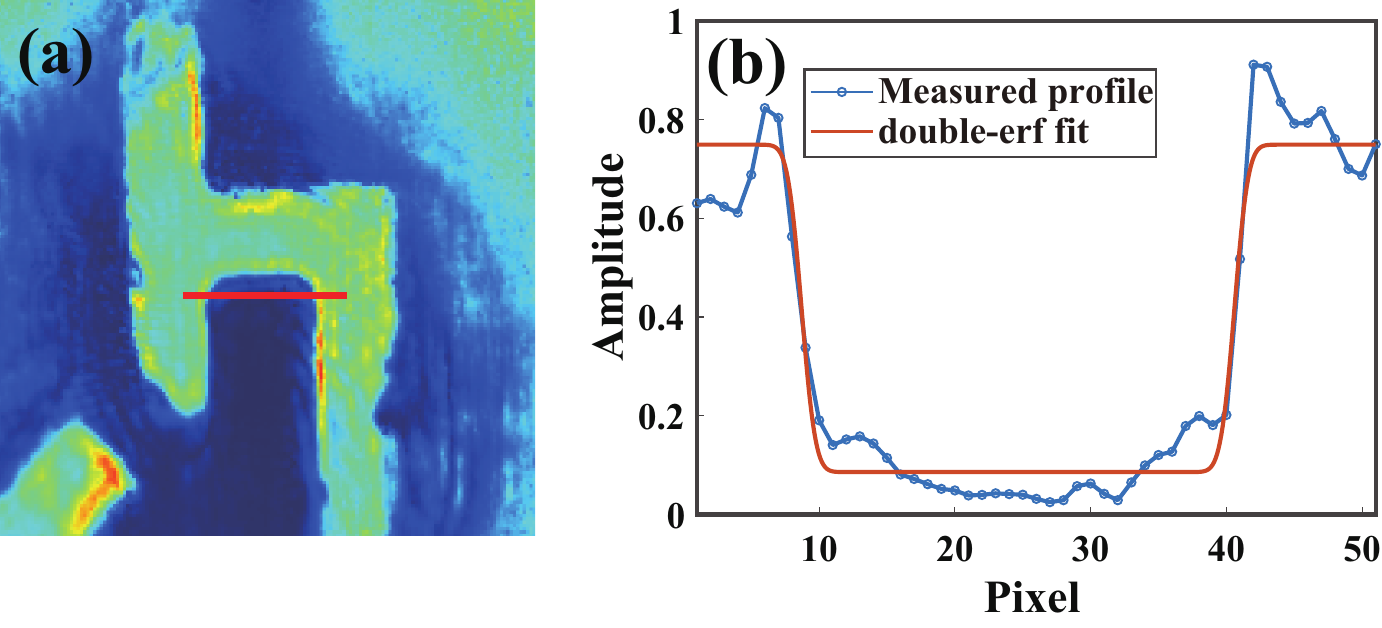}%
	\caption{\label{fig:knife} Knife-edge resolution calibration of the single-harmonic reconstruction. (a) Enlarged view of a line feature in Fig.~\ref{fig:reconstruction}(c). (b) Amplitude profile across the line region indicated in (a) and its error-function fit; the fitted line width of $31.9\pm0.3$ pixels corresponds to the designed linewidth of $30~\mu$m.}
\end{figure}

\begin{figure}
	\centering
	\includegraphics[width=\linewidth]{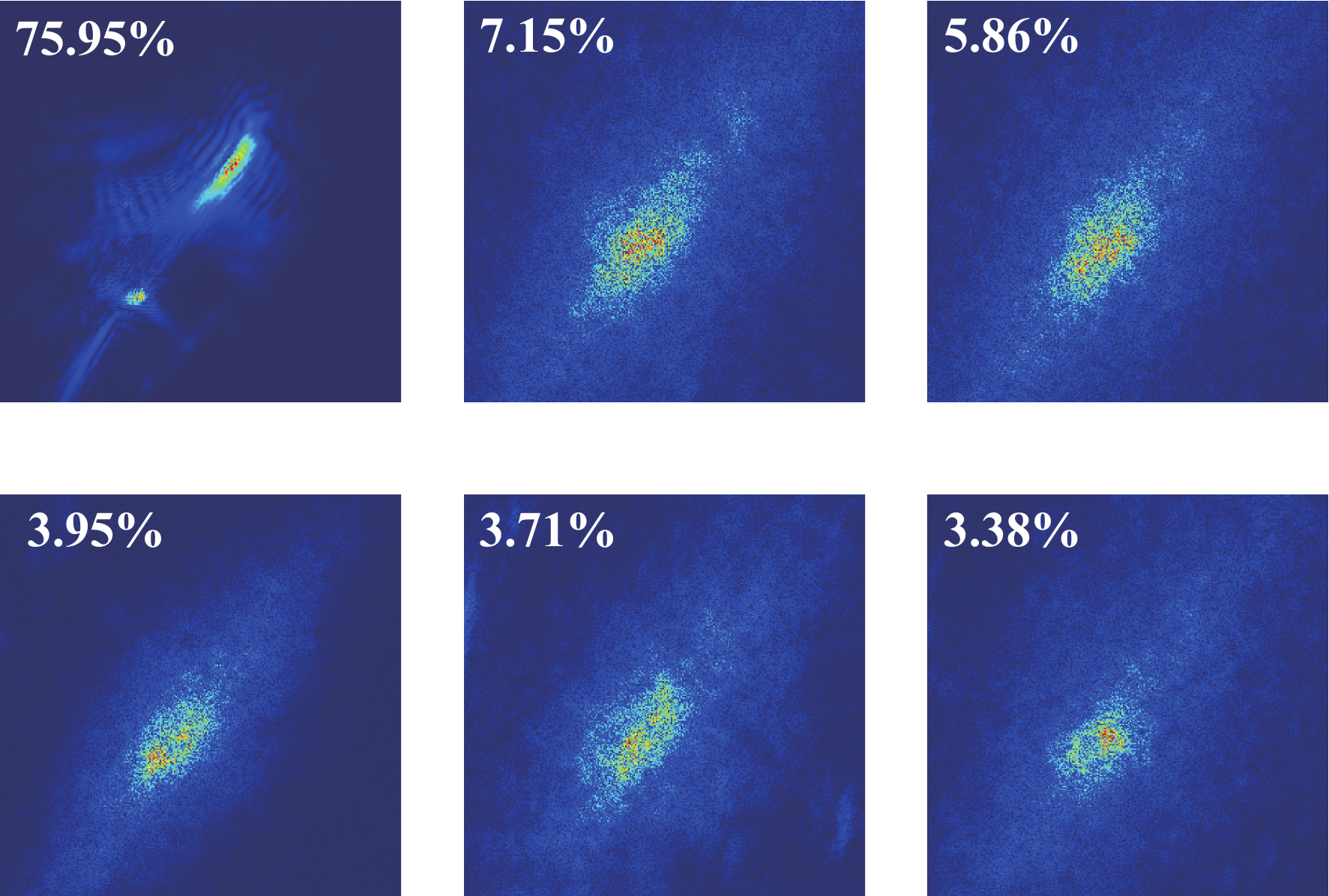}%
	\caption{\label{fig:mode} Intensities of the six reconstructed probe spatial modes in the single-harmonic reconstruction. The percentage in each panel denotes the fraction of the total probe energy carried by that mode.}
\end{figure}

In the experiment, the driving (frequency-doubled) pulses have a central wavelength of $400$~nm, and the argon pressure is about $3000$~Pa. Under these conditions, the two dominant harmonic orders are the seventh ($q=7$) and the ninth ($q=9$), corresponding to photon energies of about $21.7$~eV and $27.9$~eV and central wavelengths of about $57$~nm and $44.4$~nm, respectively. Driving the HHG process with the frequency-doubled pulses rather than the fundamental doubles the spacing between adjacent harmonic orders, so that an individual order can be isolated through the combination of the aluminum filter and power tuning. After the low-order harmonics are removed by the aluminum film, the harmonic beam illuminates the sample, and the MCP--phosphor detector records the diffraction patterns while the sample is scanned along the zigzag path.

Because the intensity of an EUV diffraction pattern spans several orders of magnitude between the low-frequency center and the high-frequency edges, each scan position was measured at three exposure times of $30$, $90$, and $270$~ms. The unsaturated regions of the individual exposures were normalized by the exposure time and stitched into a high-dynamic-range diffraction pattern. The main decay time of the P43 phosphor is about $1$~ms, much shorter than the shortest exposure of $30$~ms. Each exposure therefore integrates the steady-state response of the phosphor. For a given signal intensity, the recorded signal is proportional to the exposure time, which is the basis of the normalization and stitching described above. The exposure time can be shortened by increasing the voltage across the MCP to raise the gain. A higher gain, however, also amplifies dark counts and ion-feedback noise and increases the gain fluctuation, which degrades the signal-to-noise ratio in the weak-signal region. The exposure time and the gain must therefore be balanced according to the signal strength. The stitched patterns were then corrected by the deblurring network and reconstructed by the PIM algorithm.

\begin{figure*}
	\includegraphics[width=\linewidth]{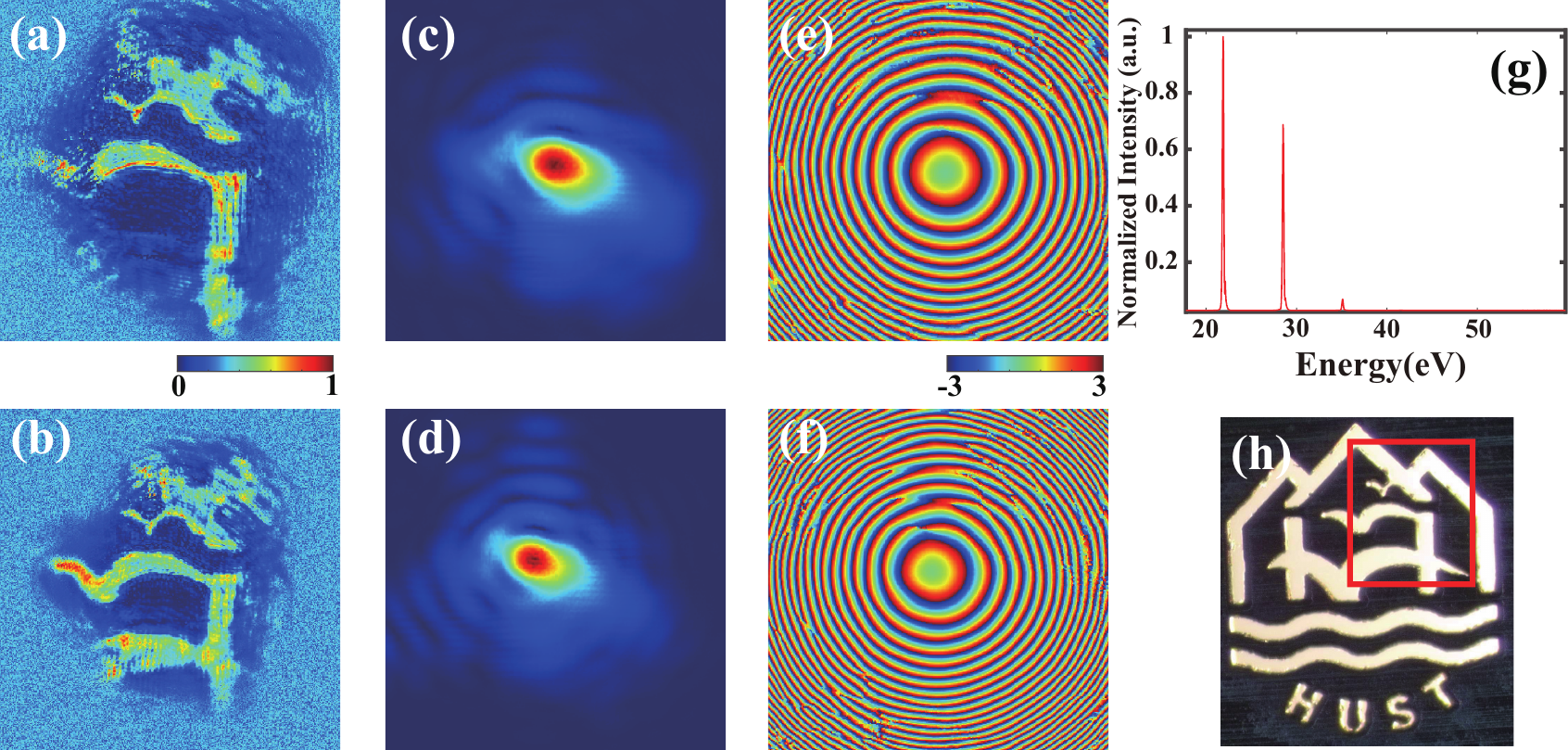}%
	\caption{\label{fig:dual} Two-harmonic ($q=7$ and $q=9$) reconstruction results. (a), (b) Reconstructed object amplitudes at $44.4$~nm ($q=9$) and $57$~nm ($q=7$), respectively. (c), (d) Corresponding reconstructed probe amplitudes. (e), (f) Corresponding reconstructed probe phases. (g) Measured harmonic spectrum. (h) The sample imaged under a metallographic microscope. The red box in (h) marks the region reconstructed in (a) and (b).}
\end{figure*}

The sample was a transmission stencil of the university emblem of Huazhong University of Science and Technology, laser-cut through a $20~\mu$m-thick aluminum foil (see Fig.~\ref{fig:dual}(h)). Two imaging experiments were carried out. In the single-harmonic experiment, the average driving power was set to about $200$~mW, under which the seventh harmonic ($q=7$, $57$~nm) dominates while the ninth harmonic is negligible. The illumination was defined by a $100~\mu$m pinhole that had been damaged in operation, so that the probe wavefront incident on the sample was highly irregular. The sample was scanned over the $10\times10$ grid with a nominal step size of $9.6~\mu$m.

As a control, the stitched patterns were first fed to the phase-retrieval algorithm without deblurring, with all other conditions unchanged. The reconstruction failed almost completely. As shown in Fig.~\ref{fig:reconstruction}(a), the reconstructed object amplitude shows only a blurred outline of the emblem, and the edges of the structure cannot be resolved. The reconstructed object phase in Fig.~\ref{fig:reconstruction}(b) is dominated by noise and shows no recognizable structure. The reconstructed probe amplitude is confined to a small bright spot (Fig.~\ref{fig:reconstruction}(e)), and the reconstructed probe phase (Fig.~\ref{fig:reconstruction}(f)) consists of strong concentric fringes and high-frequency fluctuations rather than a smooth wavefront. The recorded intensity is the convolution of the true diffraction pattern with the detector PSF, whereas the forward model used in the reconstruction assumes a sharp diffraction pattern at the detector. The algorithm cannot reconcile the overlapping constraints with the blurred intensity data, and the iteration stagnates without converging to the true object and probe. This comparison shows that the spatial blur introduced by the MCP detector during acquisition, rather than the reconstruction algorithm or the scan redundancy, is the principal obstacle to direct phase retrieval.

After correction by the deblurring network, the same data were reconstructed with the same algorithm and parameters. The corrected results are shown in Fig.~\ref{fig:reconstruction}. The reconstructed object amplitude [Fig.~\ref{fig:reconstruction}(c)] agrees well with the metallographic-microscope image of the same sample [Fig.~\ref{fig:reconstruction}(i)]; several damaged small features are circled in both panels to highlight their correspondence. The reconstructed object phase [Fig.~\ref{fig:reconstruction}(d)] shows clear contrast at the emblem edges and is approximately uniform over the foil. It changes sharply at the cut edges, consistent with a binary transmission object and with the amplitude distribution. The reconstructed probe amplitude [Fig.~\ref{fig:reconstruction}(g)] is spatially structured rather than a compact spot. The probe phase [Fig.~\ref{fig:reconstruction}(h)] shows concentric rings, which arise from diffraction at the circular aperture, together with additional distortions introduced by the damaged pinhole edge. Despite this severe wavefront irregularity, both the object and the probe are accurately recovered, which indicates that the reconstruction does not rely on an idealized or pre-characterized illumination.

The spatial resolution of the reconstruction was calibrated with a knife-edge fit on the single-harmonic result. Figure~\ref{fig:knife} shows the line feature and its amplitude profile, which is fitted with a double error function; the fitted width corresponds to the $30~\mu$m design linewidth, giving a spatial scale of $0.94~\mu$m per pixel. After correcting for the bilinear-interpolation blur introduced by the rotation, the resolution estimated from the 10--90\% width of the edge profile is $1.08~\mu$m. This value is about $19\%$ larger than the diffraction-limited resolution of $0.91~\mu$m, indicating that the residual MCP blur is no longer the dominant limitation.

To mitigate the influence of residual instabilities of the EUV source, such as fluctuations in beam pointing and pulse energy, the illumination at a single wavelength is not treated as a single deterministic field. The probe is described by its cross-spectral density, which is decomposed into a set of orthogonal spatial modes \cite{Shao2024}. The decomposition, together with the per-wavelength orthogonalization that maintains it during the iteration, is given in \ref{app:pty}. In the single-harmonic reconstruction, six spatial modes were used. Figure~\ref{fig:mode} shows the intensities of the six reconstructed probe modes; the percentage in each panel is the modal energy normalized by the total. The dominant mode carries $75.95\%$ of the total probe energy and has a well-defined structure, whereas the remaining modes are speckle-like and each carries only a few percent. This rapid decay of the mode energy indicates that the illumination retains a high degree of spatial coherence. The higher-order modes absorb the residual pointing and energy fluctuations of the EUV source, so that these instabilities are not transferred to the reconstructed object. This improves the convergence and accuracy of the reconstruction \cite{Shao2024}.

In the two-harmonic experiment, the average driving power was increased to about $400$~mW, so that both the seventh and ninth harmonics ($57$~nm and $44.4$~nm) made substantial contributions and were retained simultaneously. The sample was scanned with a smaller nominal step size of $7.84~\mu$m to provide the additional overlap redundancy required to separate the two harmonic modes. A nominally intact $50~\mu$m pinhole was used, so that the illumination closely resembled the diffraction pattern of a circular aperture. The two wavelength channels correspond to the two harmonic components, and each channel is decomposed into six spatial modes. The results shown in Fig.~\ref{fig:dual} correspond to the dominant spatial mode of each channel. The same trained network was applied to this dataset without retraining. The spatial blur, gain statistics, and noise of the detector are properties of the electron signal and are independent of the photon energy, so the correction remains applicable when both harmonics are present in the same pattern.

Figure~\ref{fig:dual} shows the reconstruction results, including the object amplitudes at $44.4$~nm and $57$~nm [Figs.~\ref{fig:dual}(a) and \ref{fig:dual}(b)], the corresponding probe amplitudes [Figs.~\ref{fig:dual}(c) and \ref{fig:dual}(d)] and probe phases [Figs.~\ref{fig:dual}(e) and \ref{fig:dual}(f)], the measured harmonic spectrum [Fig.~\ref{fig:dual}(g)], and the sample imaged under a metallographic microscope [Fig.~\ref{fig:dual}(h)]. The two object reconstructions reproduce the same stencil geometry, with coincident edge positions, which indicates that the two wavelength channels are registered consistently. The probe amplitudes are compact and nearly circular. The probe phases show regular concentric rings, as expected for the intact circular pinhole, and the additional irregularities of the single-harmonic probe are absent. The spectrum recorded with the same detector and the inserted grating is dominated by the seventh and ninth orders, and the other orders are much weaker. The ninth order is weaker than the seventh, and the object reconstruction at $44.4$~nm [Fig.~\ref{fig:dual}(a)] shows a more pronounced speckle background than that at $57$~nm [Fig.~\ref{fig:dual}(b)], consistent with the lower photon flux of the ninth harmonic. The PIM algorithm simultaneously recovers the object and the probe of both harmonic components and accurately reproduces the sample structure in each spectral channel. This confirms that CNN deblurring combined with PIM reconstruction yields high-quality multi-harmonic EUV diffraction imaging with commercially available detection hardware.

The deblurring network described in Sec.~\ref{sec:deblur-net} nevertheless has a limitation in estimating the large-scale background. It operates on the full-resolution pattern with three downsampling stages, so its effective receptive field is much smaller than the detector frame. The MCP point-spread function, however, is broad, and the modulation it imposes extends over a large fraction of the frame. A network with a limited receptive field can reproduce the local blur but cannot estimate the large-scale background reliably, so the correction of the weak outer fringes remains incomplete. This limitation arises from the representation of the input rather than from the degradation model, and it could be addressed without changing that model or the simulated training pairs.

One way to give the network a global view would be to represent the same degraded pattern at a reduced resolution. The pattern could be binned by $4\times4$ in intensity, which preserves the photon statistics and yields a view whose entire extent falls within the receptive field. The deblurring network could then be divided into two branches. The global branch would process the binned view and estimate the large-scale component of the correction, and the local branch would process the full-resolution pattern together with the upsampled output of the global branch and restore the fine fringes. The two branches would be trained jointly, so that the division of work between them is learned rather than prescribed. Because both branches would be derived from the same degraded pattern, no additional experimental data would be needed, and the correction would remain consistent with the acquisition.

\section{Conclusion}

In summary, we have presented a compact, cost-effective extreme-ultraviolet ptychographic imaging apparatus that retains high detector gain while convolutional-neural-network correction restores high spatial resolution. The method combines a physics-based simulation of the MCP detector with a residual-learning convolutional neural network to correct the detector blur computationally, so that high-spatial-resolution EUV diffraction data are obtained with modest hardware expenditure. The same MCP--phosphor detector also records the dispersed harmonic spectrum, so that no separate spectroscopic detector is required. Together with the ptychographic information multiplexing algorithm, the deblurred data enable simultaneous reconstruction of the illumination probes and of the complex transmittance of the sample at each harmonic component. The apparatus is suited to EUV imaging at the micrometer scale, wavefront diagnostics, and spectroscopy.

The significance of the method extends beyond the demonstrated apparatus and beyond CDI as an imaging modality, because it partially decouples the spatial resolution of the detector from its hardware specifications. As long as the degradation can be described by a physical model and remains stable between measurements, synthetic data can be used to train a network that recovers, after readout, the high-frequency information masked by the point-spread function, gain fluctuation, and nonlinear response. For MCP--phosphor detectors, this approach can be extended to other EUV and soft X-ray experiments, including soft X-ray microscopy, actinic inspection of EUV lithography masks, MCP-based solar and space EUV/FUV spectrographs and imagers, spatiotemporal diagnostics of laser-plasma sources, and ultrafast imaging with MCP-gated or framing cameras. It is also applicable to electron and ion imaging and spectroscopy with MCP detection, such as photoemission electron microscopy, angle-resolved photoemission spectroscopy and momentum microscopy, velocity-map imaging, electron diffraction, and imaging mass spectrometry.

Because the MCP cascade, phosphor screen, camera, and exposure conditions differ among experiments, the PSF, gain, and noise parameters in the degradation model must be recalibrated or retrained for a specific detector. The form of the degradation chain and the correction framework, however, remain unchanged. Detector degradation follows a fixed physical model and is reproducible, and can therefore be learned. Compensating hardware degradation algorithmically allows an imaging system built around a commercially available MCP to retain high gain and a large active area while approaching the diffraction limit, which provides a software-defined detector route for resource-limited EUV and soft X-ray experiments.

\appendix
\setcounter{equation}{0}
\renewcommand{\thesection}{Appendix~\Alph{section}}
\renewcommand{\theequation}{A\arabic{equation}}
\section{Ptychographic information multiplexing}\label{app:pty}

The probe and the object are indexed by the wavelength mode $w$, with central wavenumber $k_w=2\pi/\lambda_w$, and by the spatial mode $m$. Each channel has its own probe $P_{m,w}(\mathbf{r})$ and object $O_{m,w}(\mathbf{r})$. The exit wave at the $j$th scan position is
\begin{equation}
\psi_{j,m,w}(\mathbf{r}) = P_{m,w}(\mathbf{r})\,O_{m,w}(\mathbf{r}-\mathbf{r}_j),
\label{eq:exit}
\end{equation}
where $\mathbf{r}_j$ is the $j$th scan position. The exit wave is propagated to the detector plane by the Fresnel kernel,
\begin{equation}
\Psi_{j,m,w}(\mathbf{u}) = \mathcal{F}\Bigl\{\psi_{j,m,w}(\mathbf{r})\,
\exp\!\Bigl[i k_w\frac{|\mathbf{r}|^{2}}{2Z}\Bigr]\Bigr\},
\label{eq:forward}
\end{equation}
where $Z$ is the sample--detector distance and $\mathcal{F}$ denotes the Fourier transform. Because the channels are mutually incoherent, their intensities add at the detector,
\begin{equation}
I_j(\mathbf{u}) = \sum_{m,w}\bigl|\Psi_{j,m,w}(\mathbf{u})\bigr|^{2}.
\label{eq:sum}
\end{equation}
The modulus constraint distributes the measured intensity $I_j(\mathbf{u})$ among the channels in proportion to their current contributions while the phases are retained,
\begin{equation}
\Psi'_{j,m,w}(\mathbf{u}) = \sqrt{I_j(\mathbf{u})}\,
\frac{\Psi_{j,m,w}(\mathbf{u})}{\bigl(\sum_{m',w'}|\Psi_{j,m',w'}(\mathbf{u})|^{2}\bigr)^{1/2}},
\label{eq:modulus}
\end{equation}
and the corrected field is back-propagated,
\begin{equation}
\psi'_{j,m,w}(\mathbf{r}) = \mathcal{F}^{-1}\bigl\{\Psi'_{j,m,w}(\mathbf{u})\bigr\}\,
\exp\!\Bigl[-i k_w\frac{|\mathbf{r}|^{2}}{2Z}\Bigr].
\label{eq:backward}
\end{equation}
With $\Delta_{j,m,w}(\mathbf{r})=\psi'_{j,m,w}(\mathbf{r})-\psi_{j,m,w}(\mathbf{r})$ and $O_{m,w,j}(\mathbf{r})=O_{m,w}(\mathbf{r}-\mathbf{r}_j)$, the object and the probe are updated pointwise as
\begin{equation}
O_{m,w} \leftarrow O_{m,w} +
\frac{P^{*}_{m,w}\,\Delta_{j,m,w}}
{(1-a)\,|P_{m,w}|^{2}+a\,|P_{m,w}|^{2}_{\max}},
\label{eq:object}
\end{equation}
\begin{equation}
P_{m,w} \leftarrow P_{m,w} +
\frac{O^{*}_{m,w,j}\,\Delta_{j,m,w}}
{(1-b)\,|O_{m,w,j}|^{2}+b\,|O_{m,w}|^{2}_{\max}},
\label{eq:probe}
\end{equation}
where the superscript $*$ denotes the complex conjugate. The relaxation factors $a$ and $b$ lie between $0$ and $1$. The maximum-amplitude terms bound the denominators where the local amplitude approaches zero, and $a=b=1$ recovers the max-normalized form of the extended ptychographic iterative engine \cite{epie}. Cycling through all $100$ scan positions and all channels completes one iteration. For a single wavelength mode and a single spatial mode, the scheme reduces to the conventional single-mode ptychographic update.

The same physical displacement is shared by all channels, but its projection on the sampling grid depends on the wavelength, because the pixel scale at the sample plane is $\lambda_w Z/(N\Delta)$, with $N$ the array size and $\Delta$ the detector pitch. The scan positions are therefore generated separately for each wavelength mode, so that every channel refers to the same sample position despite the different pixel shifts.

The probe is also recorded with the sample removed. After each iteration, the summed far-field intensity of all channels is constrained to this measured amplitude with the distribution rule of Eq.~(\ref{eq:modulus}), applied to the probe fields instead of the exit waves.

Within a single wavelength, the illumination is described by its cross-spectral density,
\begin{equation}
J_{P,w}(\mathbf{r}_{1},\mathbf{r}_{2}) = \sum_{m=1}^{M_w} P_{m,w}(\mathbf{r}_{1})\,P_{m,w}^{*}(\mathbf{r}_{2}),
\label{eq:csd}
\end{equation}
with $\int P_{m_{1},w}^{*}(\mathbf{r})\,P_{m_{2},w}(\mathbf{r})\,d\mathbf{r}=0$ for $m_{1}\neq m_{2}$, where $M_w$ is the number of spatial modes at wavelength $w$. The modes drift out of orthogonality during the iteration and are re-orthogonalized through their Gram matrix \cite{Odstrcil2016}. Arranging the $M_w$ modes of wavelength $w$ as the rows of the matrix $P_w$, the Gram matrix is $G_w=P_wP_w^{\dagger}$, and its eigendecomposition gives
\begin{equation}
G_w = P_wP_w^{\dagger} = Q_w^{\dagger}\Lambda_w Q_w,
\label{eq:orth}
\end{equation}
\begin{equation}
\tilde{P}_w = Q_wP_w,\qquad
\tilde{P}_w\tilde{P}_w^{\dagger} = \Lambda_w.
\label{eq:orth2}
\end{equation}
where the rows of $Q_w$ are the orthonormal eigenvectors of $G_w$ and the diagonal elements of $\Lambda_w$ are the modal energies $|\tilde{P}_{m,w}|^{2}$ \cite{Shao2024}. Because $Q_w$ is unitary, the sum $\sum_m|\tilde{P}_{m,w}(\mathbf{r})|^{2}$ equals $\sum_m|P_{m,w}(\mathbf{r})|^{2}$ at every pixel, so the orthogonalization only changes the basis of the mode set and leaves the modeled detector intensity unchanged. No Gram matrix is formed across different wavelengths, because the channels are mutually incoherent and their sampling grids differ.

Isolated spikes in a probe mode are removed by comparing its amplitude with the local median. Wherever $|P_{m,w}(\mathbf{r})|$ exceeds twice the median of $|P_{m,w}|$ over a small square neighborhood, the amplitude is replaced by that median while the phase is retained.

\begin{backmatter}
\bmsection{Funding}
This work was supported by the National Key Research and Development Program of China (Grant No. 2024YFE0205800).

\bmsection{Disclosures}
The authors declare no conflicts of interest.

\bmsection{Data availability}
Data underlying the results presented in this paper are not publicly available at this time but may be obtained from the authors upon reasonable request.
\end{backmatter}

\bibliography{CNN-Corrected-EUV-Ptychography}

\end{document}